\documentclass[a4paper]{article}
\usepackage{a4,amsmath,cite}
\usepackage[dvips]{graphicx}
\usepackage{dcolumn}
\usepackage{caption}

\begin{document}

\large

\title{\bf Stable and decaying bound states on the naked Reissner-Nordstr\"om background}
\date{}

\author{V.~D.~Gladush\thanks{vgladush@gmail.com}, D.~A.~Kulikov\thanks{kulikov\_d\_a@yahoo.com}\\
\\
{\sl Theoretical Physics Department, Dniepropetrovsk National
University }\\
{\sl 72 Gagarin av., Dniepropetrovsk 49010, Ukraine} }

\maketitle

\begin{abstract}
We present a quantum mechanical study of the bound states of a
neutral scalar particle on the curved background described by the
Reissner-Nordstr\"om (RN) spacetime that corresponds to a naked
singularity. We show that there occurs both the metastable,
\textit{i.e.} decaying bound states and the stable ones. The
corresponding energy spectra are calculated in the leading WKB
approximation and the comparison with the RN black hole quasinormal
mode spectrum is made. The metastable bound states on the naked
singularity background turn out to be more long-living than the
quasinormal modes for the black hole with the same mass.
\end{abstract}

\section{Introduction}
\label{part1}

In general relativity, the probe with test particles and waves is an
efficient tool to study the properties of space–time near
gravitating mass
\cite{Chandrasekhar,Cohen,Wald,Horowitz,Ishibashi99,Pitelli}. When
using quantum test particles, one observes some features that are
unexpected from the viewpoint of classical theory. In particular, it
is well established that quantum test particles cannot form stable
stationary  bound states in the field of the Schwarzschild and
Reissner-Nordstr\"om (RN) black holes \cite{Deruelle}. Instead,
there exist decaying quasistationary states known as the quasinormal
modes (see Refs.~\cite{Kokkotas,Berti} for review).

However, if in the RN setup the gravitating mass $M$ and the
electric charge $Q$ obey the superextremality condition, $Q>M$, the
space-time geometry corresponds not to a black hole but to a naked
singularity. At the classical level, this results in a repulsive
potential barrier near the origin that prevents the absorbtion of
neutral test particles \cite{Qadir,Gladush,Pugliese}. Therefore one
may anticipate that the bound states are stable at the quantum level
as well.

The main goal of this article is to verify the above proposition and
to derive the energy spectrum of the  bound states of a quantum test
particle on the naked RN background.

Probing the naked RN singularity, one faces two key problems. The
first one concerns physical relevance. According to the cosmic
censorship conjecture by Penrose \cite{Penrose}, any singularity
originated from gravitational collapse must be hidden inside an
event horizon. Besides, since the RN singularity is electrically
charged, it may be neutralized by spontaneous pair creation provided
that charges are sufficiently high \cite{Damour}. Nevertheless, one
cannot exclude scenarios in which the RN geometry with $Q>M$ is an
exterior solution connected with a regular interior solution with
matter, for instance, the Friedmann solution in the model of
friedmon \cite{Markov}. Such models can be traced back to Dirac's
suggestion to describe leptons as charged shells \cite{Dirac}.
Recall that it is $Q/M=2.0\cdot 10^{21}$ for the electron, so that
the exterior geometry must correspond to the naked singularity.

The second problem is the non-uniqueness of the time evolution for
test particles and waves on the naked RN background
\cite{Ishibashi99,Ishibashi03,Stalker}. This means one has to
specify an additional boundary condition at the singularity to
obtain a fully unique dynamics. In the present work, we overcome
this shortcoming by employing the WKB approach, which turns out to
be insensible to behavior in the vicinity of the singularity.

The work uses a simple quantum mechanical model. We consider a
neutral scalar particle on the background geometry; the word
"particle"\; may refer equally to the test scalar field
perturbation. Both massless and massive cases are studied. This
investigation expands our previous work \cite{Kulikov}, which was
restricted to the stable bound states of a massive particle.

The plan of the article is as follows. In Section \ref{part2} we
examine how particle wave functions behave in the vicinity of the
singularity and advocate the WKB approximation. In Section
\ref{part3} the effective potential is analyzed to distinguish
between stable and decaying bound states. Section \ref{part4} is
devoted to the calculation of the energy spectra through the WKB
method. The energies obtained are then discussed and compared with
the known quasinormal mode spectra for the RN black holes in Section
\ref{part5}. Finally, Section \ref{part6} presents our conclusion.

\section{Behavior in the vicinity of a RN singularity}
\label{part2}

To start with, we establish the behavior of a particle wave function
in the vicinity of a RN singularity and study how it is accommodated
in the WKB approximation. The external RN geometry is described by
the metric (in units with $G=\hbar=c=1$)
\begin{equation}\label{RNmetric}
ds^{2} =Fdt^{2}-F^{-1}dr^{2}-r^{2}(d\theta ^{2}+\sin ^{2}{\theta }{%
d\phi }^{2})\,, \quad F =1-\frac{2 M}{r}+\frac{Q^{2}}{r^{2}}.
\end{equation}
For concreteness, we assume that $Q>0$.

The Klein-Gordon equation for a neutral scalar particle of mass $m$
on this background
\begin{equation}\label{KG}
    \frac{1}{\sqrt{-g}}\frac{\partial }{\partial x^{\mu }}(\sqrt{-g}g^{\mu \nu }
\frac{\partial \Psi }{\partial x^{\nu }})+m^{2} \Psi =0
\end{equation}
has stationary state solutions, which we decompose into the
spherical harmonics
\begin{equation}\label{Phi}
    \Psi (t,r,\theta ,{\phi })=\exp (-\frac{i\omega t}{\hbar })Y_{lm}(\theta ,{\phi })\,\frac{1}{r\sqrt{F}}u(r).
\end{equation}

Then the radial wave function satisfies the equation
\begin{equation}\label{RadEq}
    \frac{d^{2}u}{dr^{2}}+\frac{1}{F}\left[ \frac{1}{F}\left(
\omega^2-\frac{Q^2 -M^{2}}{r^{4}}\right) -m^2
-\frac{l(l+1)}{r^{2}}\right] u=0
\end{equation}
where  $l=0,1,2,...$ is the orbital quantum number. In principle,
this can be reduced to the confluent Heun equation \cite{Ronveaux},
upon substituting
$u(r)=\exp(\sqrt{m^2-\omega^2}r)(M-\sqrt{M^2-Q^2}-r)^\alpha(M+\sqrt{M^2-Q^2}-r)^\beta
y(r)$ with $\alpha$ and $\beta$ given by bulky expressions in terms
of $Q$, $M$ and $m$. The so-obtained confluent Heun equation has
singularities at $r=M\pm \sqrt{M^2-Q^2}$ (black-hole horizons) and
$r=\infty$. However, in the super-extreme case under consideration
these singularity are shifted away from the real line because
$M^2-Q^2<0$. As a consequence, the reduction to the confluent Heun
form does not simplify the problem.

It is more useful to note that the radial equation (\ref{RadEq})
holds invariant under rescaling
\begin{equation}\label{rescaling}
Q\rightarrow \alpha Q, \quad  M\rightarrow \alpha M, \quad
m\rightarrow \frac{1}{\alpha}m, \quad \omega\rightarrow
\frac{1}{\alpha}\omega, \quad R\rightarrow \alpha r
\end{equation}
and thus one of the three parameters $Q$, $M$ and $m$ can be fixed
arbitrarily.

Close to the naked singularity at $r=0$, the radial equation
simplifies
\begin{equation}\label{RadEq0}
    \frac{d^{2}u}{dr^{2}}-\frac{[1+l(l+1)]Q^2 - M^{2}}{Q^{4}}\, u=0, \qquad r\rightarrow 0.
\end{equation}
If we now adopt the leading WKB approximation, choosing the radial
wave function in the form $u(r)\propto\exp (iS(r))$, then in the
super-extreme case we have $S'^{2}(r)<0$ for small $r$. Hence, the
naked singularity is located in the classically forbidden region. As
for the bound states we are interested in, this means that the wave
function asymptotics is dominated by the term which decreases
exponentially as $r$ approaches $0$. Note that for small $r$ the
condition of the WKB-method applicability $|S''(r)/S'^{2}(r)|\ll 1$
is still satisfied.

Nevertheless, the exact behavior of the wave function at
$r\rightarrow 0$ remains undetermined. The reason is that in a pair
of linearly independent solutions to Eq.~(\ref{RadEq}) both
solutions are locally normalized at $r\rightarrow 0$ and thus none
of them can be discarded. As seen from  Eq.~(\ref{RadEq0}), such a
pair may easily be composed by taking a solution with $u(0)=0$ and
another one with $u'(r)|_{r=0}=0$. In the general case, one should
consider a linear combination of these solutions and employ the
mixed boundary condition $(u'(r)+a\,u(r))|_{r=0}=0$ with an
arbitrary real parameter $a$. Physically, the dependence of the
dynamics on this parameter indicates that the naked singularity
carries some additional degrees of freedom beyond those contained in
the metric. It is said \cite{Ishibashi99} that the RN singularity
has "hair".

Mathematically, the ambiguity in choosing the boundary condition
stems from the non-uniqueness of the self-adjoint extension to the
wave operator on the naked RN background
\cite{Ishibashi99,Ishibashi03,Stalker}. As a possible resolution,
the Friedrichs extension was suggested which implies the choice
$a=0$ and is naturally connected with the quadratic form of energy
\cite{Ishibashi03,Stalker}. It should be stressed that the
application of the WKB method will permit us to bypass this problem
at all because the method does not uses the value of the wave
function in the origin.

In the end of this Section it is worthy to compare the above picture
in terms of the areal radial coordinate $r$ with that based on the
usual tortoise coordinate $x$ defined by  $dr=Fdx$. Adjusting the
integration constant, one can always put the limit of $r\rightarrow
0$ in correspondence with $x\rightarrow 0$. Then in place of the
radial equation (\ref{RadEq}) one obtains
\begin{equation}\label{RadEqX}
    \frac{d^{2}\phi}{dx^{2}}+\left[\omega^2 -\left(m^2+ \frac{l(l+1)}{r^2}+\frac{2M}{r^3}-\frac{2Q^2}{r^4}\right)F \right] \, \phi=0
\end{equation}
where $\phi(x)=u(r)/\sqrt{F(r)}$. 
In the limit of $x\rightarrow 0$ one has $x\sim r^3/(3Q^2)$ and this
equation reduces to the Schr\"odinger-type one with the inverse
quadratic potential
\begin{equation}\label{RadEq1}
    \frac{d^{2}\phi}{dx^{2}}+\frac{2}{9x^{2}}\, \phi=0.
\end{equation}
Then the textbook analysis  \cite{Landau} shows that the particle
cannot fall on the origin because the numerical coefficient of the
potential in Eq.~(\ref{RadEq1}) does not exceed its critical value
1/4. This is in agreement with our previous conclusion that the
naked singularity is located in the classically forbidden region.

However, Eqs.~(\ref{RadEqX}) and (\ref{RadEq1}) contain the
singularity at $x=0$ such that the condition of the WKB-method
applicability breaks down for small $x$. Therefore, to develop the
WKB approximation in the subsequent sections, we employ the initial
radial equation (\ref{RadEq}) in terms of the areal radial
coordinate $r$.

\section{Effective potential}
\label{part3}

Now we shall establish possible types of the particle bound states
on the naked RN background. To that end, let us analyze the shape of
the effective potential $V(r)$ introduced upon rewriting the radial
equation (\ref{RadEq}) in shorthand notations
\begin{equation}\label{SecEq}
    \frac{d^{2}u}{dr^{2}}+\frac{1}{F^{2}}\left(\omega^{2}-V(r)\right)u=0,
\end{equation}
where
\begin{eqnarray}\label{Veff}
    V(r)&=&\left(m^2+\frac{l(l+1)}{r^{2}}\right)F+\frac{Q^{2}-M^2}{r^{4}} \nonumber \\
    &=&{m}^{2}-{\frac {2M{m}^{2}}{r}}+{\frac {{Q}^{2}{m}^{2}+\Lambda}{{r}^{2
}}}-{\frac {2\Lambda M}{{r}^{3}}}+{\frac
{{Q}^{2}\Lambda-{M}^{2}+{Q}^{ 2}}{{r}^{4}}}
\end{eqnarray}
and we designated $\Lambda=l(l+1)$. Note that  $V(r)$ is
positive-defined  because $F>0$ for all positive $r$.

For a given particle energy $\omega$, the type of the bound state is
regulated by the pattern of classically allowed and forbidden
regions. From Eq.~(\ref{SecEq}) one concludes that the classically
allowed (forbidden) region is to be defined as a region in which the
condition $\omega^{2}-V(r)>0$ ($\omega^{2}-V(r)<0$) holds.

Instead of solving those inequalities explicitly, we are looking for
local extrema of $V(r)$. If $V(r)$ has a minimum at $r=r_{min}$ and
no other extrema, then for $\omega$ close enough to
$\sqrt{V(r_{min})}$ the classically allowed region consists of a
single line segment and there stable bound states may exist. In this
case we call $V(r)$ the single-well potential. If $V(r)$ has two or
more extrema, then for certain $\omega$ the classically allowed
region must be split into two subregions separated by the
classically forbidden region. Because of the tunneling between the
subregions, the bound states are now metastable, \textit{i.e.}
decaying. The corresponding potential will be referred to as the
barrier-shaped one.

We are now in the position to deduce what type of the bound states
occurs depending on the values of the particle mass $m$ and the
singularity parameters $Q$ and $M$. Let us first study the simpler
case of the massless particle and then turn to the massive case.

\subsection{Case of $m=0$}

In this case the effective potential simplifies
\begin{eqnarray}\label{Veffm0}
    V(r)={\frac {\Lambda}{{r}^{2
}}}-{\frac {2\Lambda M}{{r}^{3}}}+{\frac
{{Q}^{2}\Lambda-{M}^{2}+{Q}^{ 2}}{{r}^{4}}}.
\end{eqnarray}
Notice that for the S-waves ($l=0 \Rightarrow \Lambda=l(l+1)=0$) it
is monotonically decreasing and thus only scattering states are
present in the spectrum.

For $l\neq 0$ the formula
\begin{eqnarray}\label{crit-m0}
\left(\frac{Q}{M}\right)_{cr}^2 = \frac{8+9\Lambda}{8+8\Lambda}
\end{eqnarray}
defines a critical charge-to-mass ratio such that for
$Q/M>(Q/M)_{cr}$ the effective potential $V(r)$ has no extrema. If
$1<Q/M<(Q/M)_{cr}$, there must be one local minimum and one local
maximum, so that the potential is barrier-shaped and there emerge
the metastable bound states. Noticeably, in the limiting case of the
extreme black hole, $Q/M\rightarrow 1$, the local minimum transforms
in an event horizon.

From Eq.~(\ref{crit-m0}), one sees that the value $(Q/M)_{cr}$
increases with the growth of $\Lambda$, though not exceeding
$\sqrt{9/8}\approx 1.0607$. Thus the metastable bound states may
only exist in the narrow range $1<Q/M<(Q/M)_{cr}$. The plots of
$V(r)$ for $m=0$, obtained with the values $Q/M=1.01$ and $1.1$
inside and outside this range, respectively, are shown in Fig.~1.

\begin{figure}[h]
\begin{tabular}{cc}
 \includegraphics[scale=0.3]{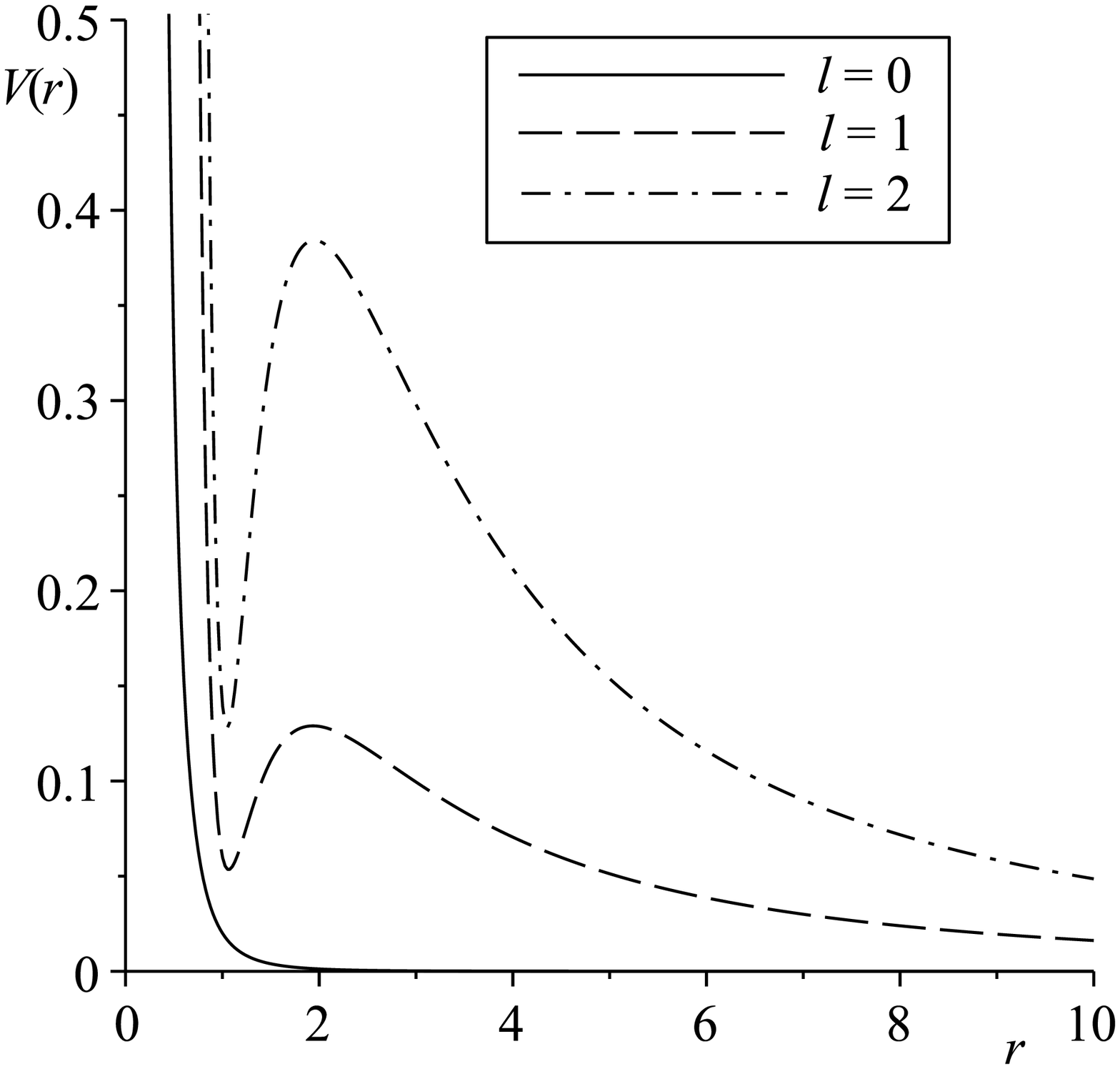} &  \includegraphics[scale=0.3]{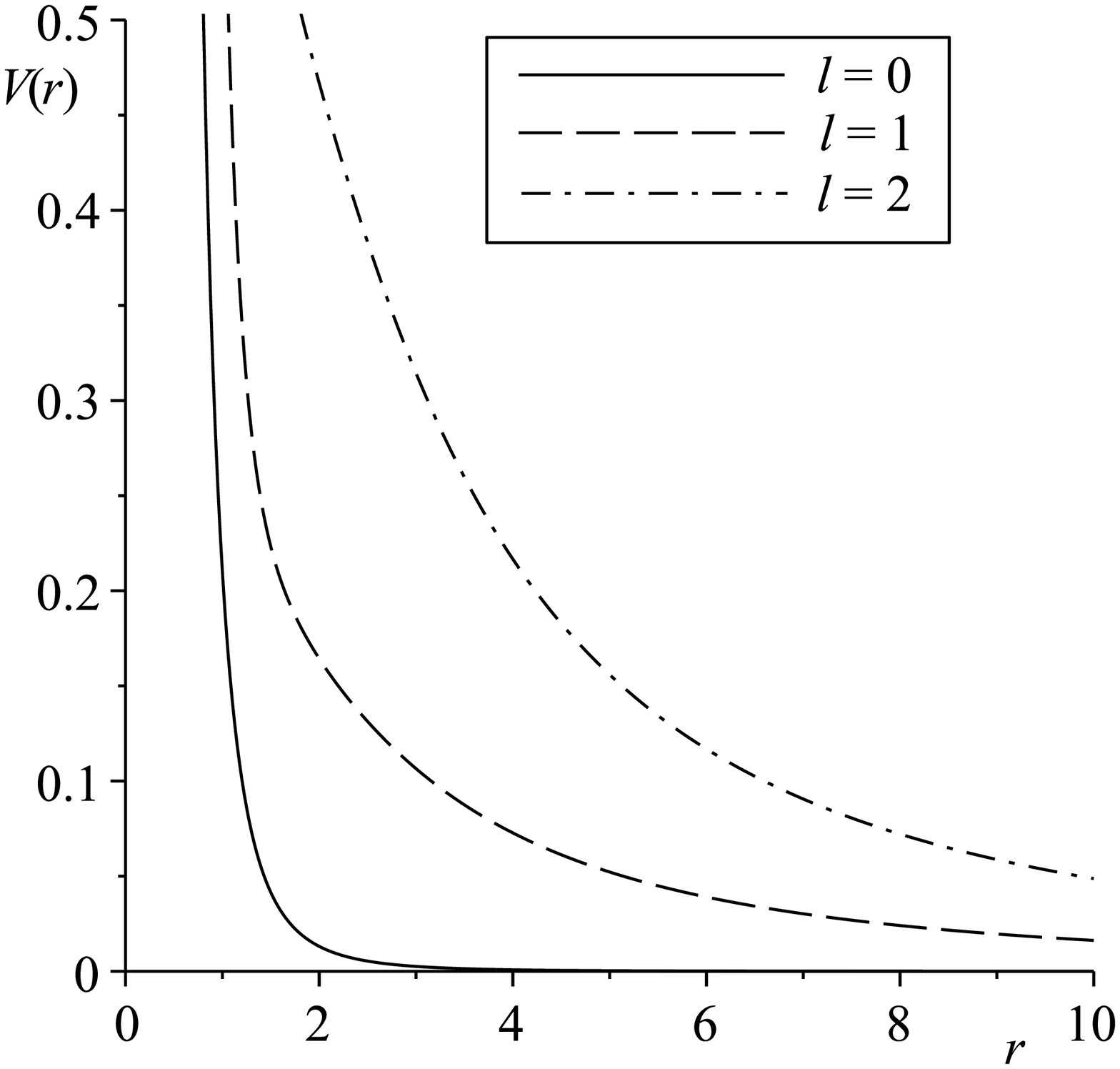} \\
  (a) & (b) \\
\end{tabular}
\label{fig1} \small \caption{ Effective potential  $V(r)$ for the
particle with $m=0$  on the background with (a) $M=1$, $Q=1.01$, (b)
$M=1$, $Q=1.1$. Different lines refer to the orbital numbers $l= 0,
1, 2$. \hfill}
\end{figure}

\subsection{Case of $m\neq 0$}

We shall perform the same analysis as in the previous subsection but
for the massive particle. For brevity, we set $M=1$ in this
subsection; the formulae with arbitrary $M$ can readily be obtained
by rescaling (\ref{rescaling}).

To study the shape of $V(r)$, we find its extrema that amounts to
solving the cubic equation
\begin{equation}\label{cubic1}
\frac{dV}{dr}=\frac{2m^2}{r^5}(r^{3}+ar^{2}+br+c)=0
\end{equation}
with the coefficients
\begin{equation}\label{cubic-coef}
a=-Q^{2}-\frac{\Lambda}{m^2},\quad b=\frac{3\Lambda}{m^2},\quad
c=-\frac{2[(1+\Lambda)Q^{2}-1]}{m^2}.
\end{equation}
The number of real roots to this cubic equation is controlled by the
sign of its determinant defined by
\begin{equation}\label{cubic-D}
D=\left(\frac{p}{3}\right)^3+\left(\frac{q}{2}\right)^2
\end{equation}
where
\begin{equation}\label{cubic-pq}
p=-\frac{a^2}{3}+b,\quad
q=2\left(\frac{a}{3}\right)^3-\frac{ab}{3}+c.
\end{equation}
There must be one, two and three real roots when $D>0$, $D=0$ and
$D<0$ respectively. Since the signs of the coefficients alternate,
all these roots are positive and thus all local extrema of $V(r)$
are of physical meaning.

It is convenient to rewrite $D$ in terms of the deviation from the
extreme charge value $\delta=Q^2-1$
\begin{eqnarray}\label{Discr}
D&=&\frac{1}{108m^8}\left[ \left( 8\,\delta-1 \right) {\Lambda}^{4}+
\left(
-78\,{m}^{2}\delta+6\,{m}^{2}+24\,{m}^{2}{\delta}^{2}+8\,\delta
\right) {\Lambda}^{3} \right. \nonumber \\
&+& \left(
24\,{m}^{4}{\delta}^{3}+15\,{m}^{4}-84\,{m}^{2}\delta+24\,{m}^
{2}{\delta}^{2}+63\,{m}^{4}{\delta}^{2}+54\,{m}^{4}\delta \right)
{ \Lambda}^{2} \nonumber \\
&+& \left( 8\,{m}^{6}{\delta}^{4}+32\,{m}^{6}\delta+48\,{m}^
{6}{\delta}^{2}+32\,{m}^{6}{\delta}^{3}+8\,{m}^{6}+24\,{m}^{4}{\delta}
^{3}+156\,{m}^{4}{\delta}^{2}\right.
 \nonumber \\
 &+&\left.\left. 132\,{m}^{4}\delta \right)\Lambda+8\,{m
}^{6}\delta+24\,{m}^{6}{\delta}^{2}+24\,{m}^{6}{\delta}^{3}+8\,{m}^{6}
{\delta}^{4}+108\,{m}^{4}{\delta}^{2}\right].
\end{eqnarray}

For the super-extreme charge under consideration, we have
$\delta>0$. Then it becomes evident from (\ref{Discr}) that the
condition $D>0$ always holds true for the S-states ($\Lambda=0$)
and, hence, the corresponding effective potential has one local
extremum (minimum), thereby being single-well. Thus, in contrast to
the case of $m=0$, the particle with $m\neq 0$ may form stable bound
S-states.

As an illustration, in Fig.~2 we offer plots of $V(r)$  obtained
with $m=0.1$, using  $Q/M=1.01$ and $1.1$. In these plots the
potential well where the bound states emerge lies below the rest
energy level $\omega^2=m^2$ depicted by the dotted line.

\begin{figure}[h]
\begin{tabular}{cc}
 \includegraphics[scale=0.3]{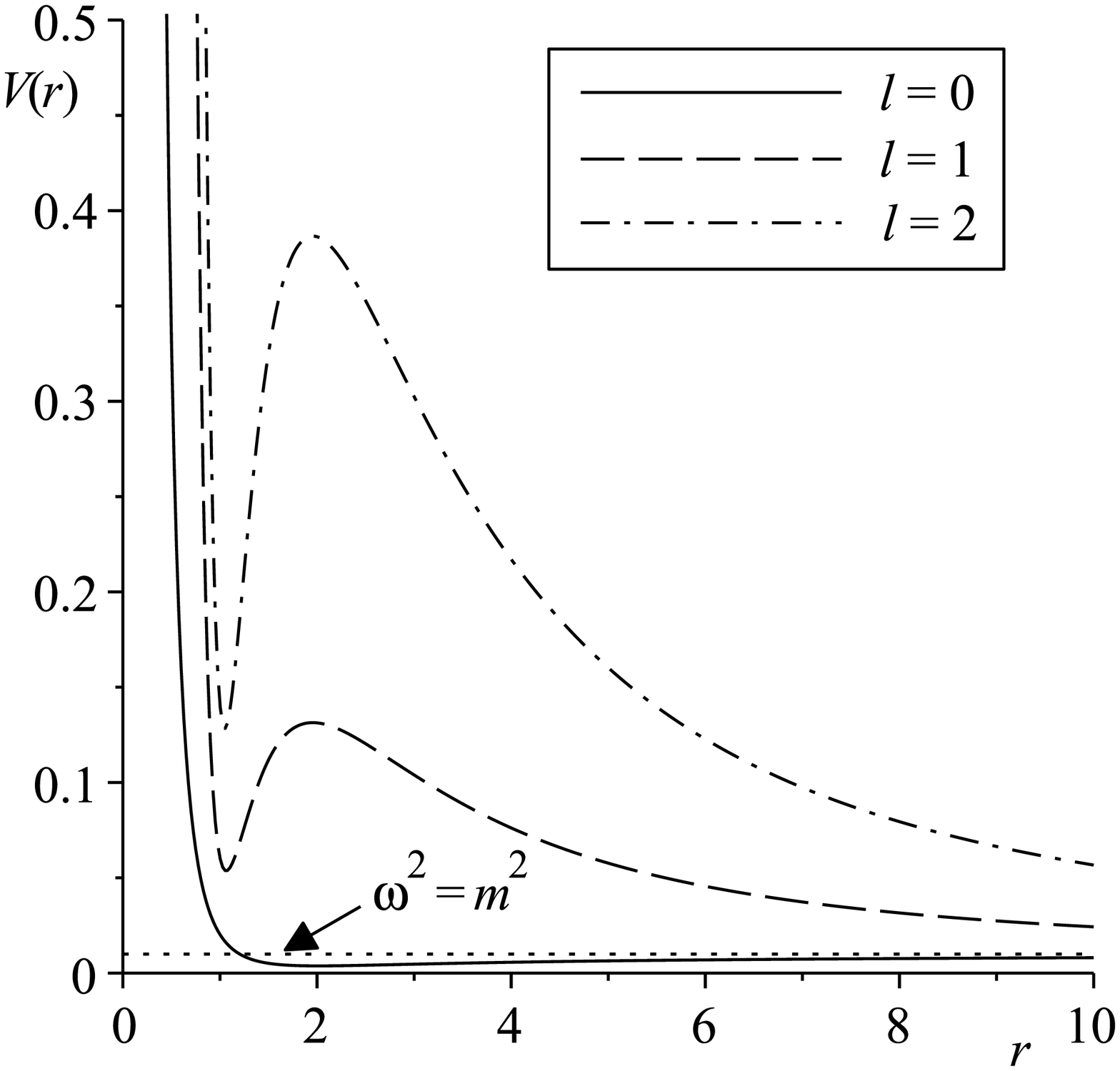} &  \includegraphics[scale=0.3]{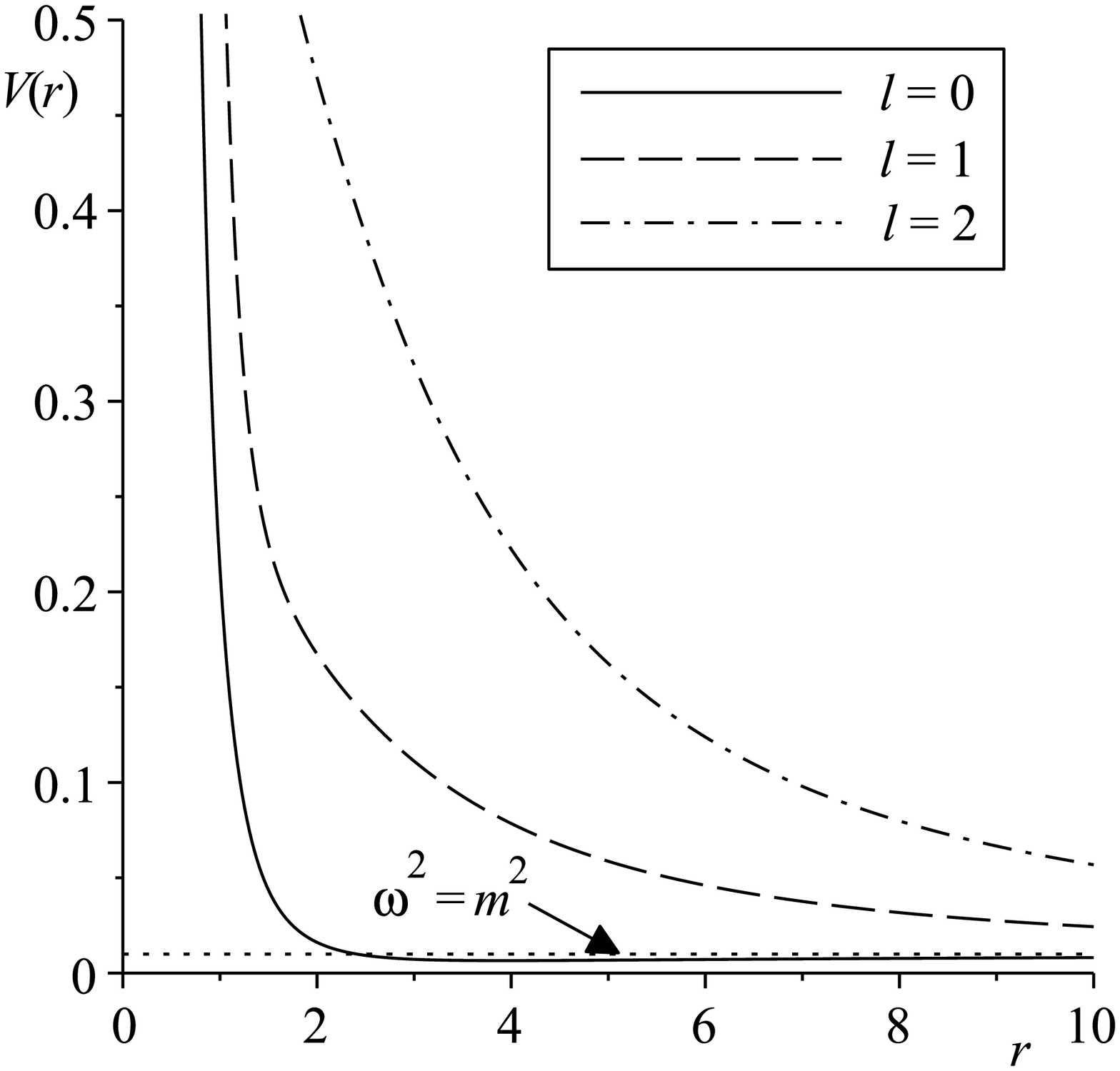} \\
  (a) & (b) \\
\end{tabular}
\label{fig2} \small \caption{The same as Fig.~1 but for the particle
with $m=0.1$. \hfill}
\end{figure}

As for the spectrum with $l\neq 0$, we have learnt that in the case
of $m=0$ it contains the metastable bound states provided the
charge-to-mass ratio is close enough to its extreme value $Q/M=1$.
We now argue that in the case of $m\neq 0$ the situation is
essentially the same. Indeed, for small $\delta$ the asymptotics of
$D$ is given by
\begin{equation}\label{cubic-D1}
D=\,{\frac {\Lambda\, \left( 8\,{m}^{2}-\Lambda \right)
 \left( \Lambda+{m}^{2} \right) ^{2}}{108{m}^{8}}+O(\delta)},
\end{equation}
so that $D<0$ if $\Lambda>8m^2$.  In practice, the last condition
holds for all $l\neq 0$, because  $m\ll M=1$ (the particle mass is
supposed to be much less than the central object mass). The
inequality $D<0$ means that $dV/dr$ has three zeros and, as a
consequence, $V(r)$ is the barrier-shaped potential with two minima
and one maximum. This subcase may include the metastable bound
states because one of the minima lies above the level of
$\omega^2=m^2$ (see Fig.~2a).

On the other hand, if $Q/M$ is far from being extreme, the
metastable bound states are excluded. In the case of $m=0$, the
critical value (\ref{crit-m0}) that provides the necessary and
sufficient condition for their exclusion was derived. Now, for
$m\neq 0$, we are able to deduce the sufficient but not necessary
condition. To do this, we should establish for which values of
$\delta$ the condition $D>0$ holds, which guarantees that the
effective potential is single-well.

Upon demanding that the coefficients in front of powers of $\Lambda$
in (\ref{Discr}) be positive, one obtains several restrictions. The
strongest one comes from the coefficient of the ${\Lambda}^{2}$-term
which becomes positive-defined  when $\delta>7/2$. Thus we conclude
that the condition $D>0$ holds and, hence, the effective potential
is single-well and the metastable bound states are excluded,
provided
\begin{equation}\label{Qcrit}
   \frac{Q^2-M^2}{M^2} >{\frac {7}{2}}, \quad \frac{Q}{M}
>\frac {3\sqrt {2}}{2}\approx 2.121...
\end{equation}
It should be stressed that this is not an exact critical value and,
in fact, $D>0$ may hold at lesser values of $Q/M$ when $l$ is small
enough. For example, if $Q/M=1.1<2.121$ the effective potential
turns out to be single-well for $l=0,1,2$. In the case of $l=0$ it
can be seen from Fig.~2b whereas for $l=1,2$ the bottom of the well
is located in the region of large $r$ outside the plot.

Let us summarize our findings. For $l=0$, the spectrum contains only
the scattering states, if the particle is massless, and in addition
the stable bound states, if $m\neq 0$. For $l\neq 0$, the metastable
bound states emerge if the central object charge-to-mass ratio is
close to its extreme value $Q/M=1$. When $Q/M$ exceeds a certain
critical value, the metastable bound states disappear and one
observes the scattering states and the stable bound states.

\section{Energy spectrum}
\label{part4}

In this Section we derive the energy spectrum of the particle (=
test scalar field perturbation) in the leading WKB approximation.

Since the background geometry is naked rather than the black-hole
one, our calculation will differ essentially from typical
computations of the black-hole quasinormal modes. The latter imply
that the solutions to the wave equation go as plane waves close to
the horizon. Instead, in our case the solutions have to decrease
exponentially deep into the classically forbidden region in the
vicinity of the origin.

Actually, the barrier-shaped potentials that lead to the metastable
bound states in our treatment (see Figs.~1a and 2a) resemble those
of Gamow's theory for $\alpha$-decay (Fig.~3). In turn, the
single-well potential we obtained with $m\neq 0$ and $l=0$ resembles
the attractive Coulomb potential everywhere except for a narrow
region close to the origin. Therefore we shall adopt the ordinary
flat-space WKB formulae \cite{Landau} with the Langer modification
for the centrifugal term, $\Lambda=l(l+1)\rightarrow(l+1/2)^2$,
which is usual in the Coulomb-like problems \cite{Heading,Berry}.

\begin{figure}[th]
\centerline{\includegraphics[scale=0.22]{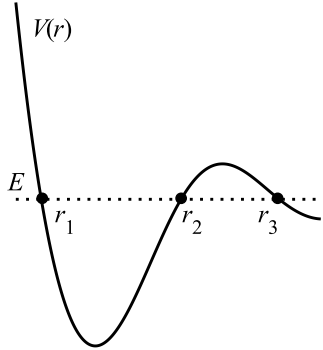}} \label{fig3}
\small \caption{ Potential in the $\alpha$-decay theory. \hfill}
\end{figure}

\subsection{Decaying bound states}

Let us start with the metastable bound states. From the wave
equation (\ref{SecEq}), it can be seen that the quasiclassical
momenta $k(r)$ and $\kappa(r)$ for the classically allowed and
forbidden regions respectively should be defined by
\begin{equation}\label{quasimom}
    k(r)=\frac{1}{F}\sqrt{\omega^2-V(r)}, \qquad
    \kappa(r)=\frac{1}{F}\sqrt{V(r)-\omega^2}.
\end{equation}
For the metastable bound states the energy
$\omega=\omega_R+i\omega_I$ is complex. In the lowest WKB
approximation its real part is determined by the Bohr-Sommerfeld
rule \cite{Landau}
\begin{equation}\label{BSrule}
\int_{r_1}^{r_2} \frac{1}{F}\sqrt{\omega_R^2-V(r)}dr=\pi (n+1/2)
\end{equation}
where $r_{1,2}$ are the turning points at the ends of the potential
well (see Fig.~3) and $n$ denotes the excitation number.

Further, we derive a Gamow-type formula for the imaginary part of
$\omega$. Our derivation follows the conventional procedure
\cite{Mur}. First, multiplying the wave equation (\ref{SecEq}) by
the complex conjugate wave function, we compose the expression
$u''\bar{u}-\bar{u}''u = -(\omega^2-\bar{\omega}^2)F^{-2}|u|^2$ and
then integrate it to obtain
\begin{equation}\label{eq-conj}
\left.\left(u'\bar{u}-\bar{u}'u\right)\right|_{r\rightarrow\infty} =
-4i\omega_R \omega_I\int_{0}^{\infty}F^{-2} |u|^2 dr.
\end{equation}
Here we took into account that $(u'\bar{u}-\bar{u}'u)$ vanishes at
$r=0$ because $u$ must satisfy the boundary condition
$u'(0)+au(0)=0$ with real $a$ in order that the Klein-Gordon
operator be symmetric.

Next, we resort to the well-known WKB formulae for the wave function
asymptotics in the classically allowed region \cite{Landau}
\begin{equation}\label{WKB-wf1}
  u\sim\frac{C_1}{2\sqrt{k(r)}}\exp\left(i\int_{r_3}^{r}k(r)dr-i\pi/4
\right),\quad r>r_3,
\end{equation}
\begin{equation}\label{WKB-wf2}
  u\sim\frac{C_2}{\sqrt{k(r)}}\cos\left(\int_{r_1}^{r}k(r)dr-\pi/4
\right),\quad r_1<r<r_2.
\end{equation}

Using the first of these formulae, the left-hand side of
(\ref{eq-conj}) is estimated to be equal to $i|C_1|^2/2$. The main
contribution to the right-hand side of (\ref{eq-conj}) comes from
the region $r_1<r<r_2$. Approximating the squared cosine in this
region by $1/2$, we rewrite (\ref{eq-conj}) as
\[
\frac{i|C_1|^2}{2} = -4i\omega_R
\omega_I|C_2|^2\int_{r_1}^{r_2}\frac{dr}{2F^2 k(r)}.
\]

Applying the connection formula \cite{Landau}
$C_2=C_1\exp(\int_{r_2}^{r_3}\kappa(r)dr)$, we end up with the
requisite expression for the imaginary part of $\omega$
\begin{equation}\label{Width}
\omega_I={\displaystyle -\exp\left(-2 \int_{r_2}^{r_3} \kappa(r)dr
\right)}\left[\displaystyle 4\omega_R\int_{r_1}^{r_2} \frac{dr}{F^2
k(r)} \right]^{-1}.
\end{equation}
This formula as well as the WKB asymptotics (\ref{WKB-wf1}) and
(\ref{WKB-wf2}) is valid if the barrier is nearly impenetrable. It
means that the calculated value of $\omega_I$, which determines the
decay rate, has to obey the condition $|\omega_I|\ll \omega_R$.

\subsection{Stable bound states}

Now let us turn to the stable bound states occurring in the case of
the massive particle. If $m\neq 0$, the effective potential
(\ref{Veff}) has the Coulomb tail ($V\propto -1/r$), so that a
Balmer-type approximate formula for the real bound-state energies
$\omega=\omega_R$ can be obtained.

To do this, we first compare the wave equation (\ref{SecEq}) for our
effective potential and the ordinary Schr\"odinger equation for the
Coulomb potential
\begin{equation}\label{CoulEq}
    \frac{d^{2}\psi}{dr^{2}}+2m\left(E+\frac{Ze^2}{r}-\frac{l(l+1)}{2mr^2} \right)\psi=0.
\end{equation}
Since at large $r$ the factor $F$ of the RN metrics (\ref{RNmetric})
tends to 1, it becomes evident that the term $2Mm^2/r$ in
Eq.~(\ref{SecEq}) corresponds to $2mZe^2/r$ in Eq.~(\ref{CoulEq}).
Then we can define the characteristic radius for our system,
$r_B=1/(m^2 M)$, corresponding to the Bohr radius, $1/(mZe^2)$, for
Eq.~(\ref{CoulEq}). Introducing now the dimensionless coordinate
$\rho=r/r_B$, the effective potential (\ref{Veff}) with the Langer
modification is rewritten as
\begin{eqnarray}\label{VeffX}
    V(\rho)&=&{m}^{2}-{\frac {2M^2{m}^{4}}{\rho}}+{\frac {M^2{m}^{4}[{Q}^{2}{m}^{2}+(l+1/2)^2]}{{\rho}^{2
}}}-{\frac {2(l+1/2)^2 M^4 m^6}{{\rho}^{3}}} \nonumber \\
&+&{\frac {M^4 m^8\{{Q}^{2}[1+(l+1/2)^2]-{M}^{2}\}}{{\rho}^{4}}}.
\end{eqnarray}
The last two terms in this expression contain higher powers of the
particle mass $m$ which is much less than the central object mass
$M$. Hence, we may neglect these terms provided that $M$ is not much
larger than the Planck mass (equal to 1 in our units). Thus we
obtain the truncated potential
\begin{eqnarray}\label{VeffTr}
    V_{trunc}(\rho)&=&{m}^{2}-{\frac {2M^2{m}^{4}}{\rho}}+{\frac {M^2{m}^{4}[{Q}^{2}{m}^{2}+(l+1/2)^2]}{{\rho}^{2
}}}
\end{eqnarray}
serving as a good approximation to $V(\rho)$ in the classically
allowed region. Using the same reasoning, we approximate $F=1-2M^2
m^2/\rho+M^2 Q^2 m^4/\rho^2$ by $1$.

Within the above approximation, the Bohr-Sommerfeld integral
(\ref{BSrule}) reduces to
\begin{equation}\label{BSruleTr}
\int_{\rho_1}^{\rho_2} \sqrt{\omega^2-{m}^{2}+{\frac
{2M^2{m}^{4}}{\rho}}-{\frac
{M^2{m}^{4}[{Q}^{2}{m}^{2}+(l+1/2)^2]}{{\rho}^{2 }}}}
\frac{d\rho}{m^2 M}=\pi (n+1/2)
\end{equation}
where turning points $\rho_{1,2}$ are zeros of the expression under
the square root. This integral is easily calculated, by applying the
formula
\begin{equation}\label{BSruleEval}
\int_{\rho_1}^{\rho_2}
\frac{\sqrt{(\rho-\rho_1)(\rho_2-\rho)}}{\rho}\, d\rho=\pi
\left(\frac{\rho_1+\rho_2}{2}-\sqrt{\rho_1 \rho_2} \right).
\end{equation}

As a result, we get the explicit expression for the stable
bound-state energies
\begin{equation}\label{WKBenergies}
  \omega=m\left[ 1-\frac{m^2 M^2}{\left(n+1/2+\sqrt{(l+1/2)^2+Q^2 m^2}\right)^2}\right]^{1/2}.
\end{equation}
It has essentially the same structure as the Balmer formula in the
Coulomb problem, assuming that the values of the quantum numbers are
high, $n, l\gg 1$, as usual in the WKB method. The only difference
is that the coupling constant is now given by the product of masses,
but not charges. Note that the quantization procedure that starts
from the classical particle Hamiltonian has lead us to the same
expression \cite{Kulikov}.

\section{Numerical results}
\label{part5}

First we examine the spectrum of the metastable bound states for the
massless neutral scalar particle in the field of naked RN
singularity. It should be noticed that since the depth and the width
of the corresponding potential well are finite, the number of the
metastable bound states is limited. The energies of all the existing
states with $l=1, 2, 3$ calculated according to the WKB formulae
(\ref{BSrule}) and (\ref{Width}) with $Q=1.01$, $M=1$ are presented
in Table 1. That the states with larger excitation numbers $n$ do
not exist was fixed by observing that the Bohr-Sommerfeld integral
(\ref{BSrule}) cannot be saturated to its value $\pi (n+1/2)$ even
at the maximal allowed bound-state energy $\omega^2$ equal to the
peak of the effective potential. As seen from Table 1, for higher
$l$ the number of the existing bound states and their half-lives
increase as the peak of the effective potential grows with $l$ (see
Fig.~1a).

\begin{table}[pht]
\large \caption{Energies of the metastable bound states on the naked
RN background calculated with $Q=1.01$,  $M=1$ and $m=0$.}
\begin{center}
{\begin{tabular}{cccc} \hline
& \multicolumn{3}{c}{ $l=1$} \\
\cline{2-4}
  $\phantom{\displaystyle{\sum}}$ & $\omega_{WKB}$ & $\sqrt{V_{min}}$ & $\sqrt{V_{max}}$ \\
\hline
 $n=0$ & $0.294636-i\,0.263\times 10^{-3}$ & 0.241152 & 0.380747 \\
\hline
& \multicolumn{3}{c}{ $l=2$} \\
\cline{2-4}
$\phantom{\displaystyle{\sum}}$   & $\omega_{WKB}$ & $\sqrt{V_{min}}$ & $\sqrt{V_{max}}$ \\
\hline
 $n=0$ & $0.423194-i\,0.164\times 10^{-5}$ & 0.364334 & 0.632606 \\
 $n=1$ & $0.537046-i\,0.227\times 10^{-3}$ &  &  \\
 \hline
 & \multicolumn{3}{c}{ $l=3$} \\
\cline{2-4}
$\phantom{\displaystyle{\sum}}$   & $\omega_{WKB}$ & $\sqrt{V_{min}}$ & $\sqrt{V_{max}}$ \\
\hline
 $n=0$ & $0.555002-i\,0.969\times 10^{-10}$ & 0.494094 & 0.884899 \\
 $n=1$ & $0.673735-i\,0.236\times 10^{-5}$ &  &  \\
 $n=2$ & $0.786160-i\,0.211\times 10^{-3}$ &  &  \\
 $n=3$ & $0.884745-i\,0.523\times 10^{-2}$ &  &  \\
\hline
\end{tabular}}\end{center}
\label{ta1}
\end{table}

\begin{table}[pht]
\large \caption{Quasinormal energies of the extreme RN black hole
with $Q=M=1$ and $m=0$ taken from Ref.~\cite{Onozawa}}
\begin{center}
{\begin{tabular}{cccc} \hline
& \multicolumn{3}{c}{ $l=1$} \\
\cline{2-4}
$\phantom{\displaystyle{\sum}}$  & $\omega_{WKB}$ & $\sqrt{V_{min}}$ & $\sqrt{V_{max}}$ \\
\hline
 $n=0$ & $0.37764-i\,0.08936$ & 0 & 0.375  \\
 $n=1$ & $0.34392-i\,0.27828$ &  & \\
 $n=2$ & $0.29661-i\,0.48145$ &  &  \\
\hline
& \multicolumn{3}{c}{ $l=2$} \\
\cline{2-4}
$\phantom{\displaystyle{\sum}}$  & $\omega_{WKB}$ & $\sqrt{V_{min}}$ & $\sqrt{V_{max}}$ \\
\hline
 $n=0$ & $0.62609-i\,0.08873$ & 0 & 0.625 \\
 $n=1$ & $0.60677-i\,0.26944$ &  &  \\
 $n=2$ & $0.57254-i\,0.45750$ &  &  \\
\hline
\end{tabular}}\end{center}
\label{ta2}
\end{table}

For comparison, in Table 2 we list the first three quasinormal
energies (frequencies) of the scalar field on the extreme RN
black-hole background with $Q=M=1$ that were computed in
Ref.~\cite{Onozawa}.

The striking difference between the results in Tables 1 and 2 can be
ascribed to the fact that the event horizon is absent in the naked
singularity case and thus the metastable bound states and the
quasinormal modes are of different nature. As was discussed in the
previous Section, the setup for the naked RN bound states resembles
that of Gamow's theory for $\alpha$-decay. In this setup the most
longliving states are concentrated near the bottom of the effective
potential well $\omega^2\simeq \mathrm{min}[V(r)]$. This can be
readily checked by comparing the calculated energies $\omega$ with
the values $V_{\mathrm{min}}= \mathrm{min}[V(r)]$ which are also
shown in Table 1. In turn, the black-hole quasinormal modes are,
roughly speaking, the scattering resonances travelling both to $r=0$
and to $r=\infty$. Thus they have much smaller half-lives and
survive for energies close to the peak of the effective potential
$\omega^2\simeq \mathrm{max}[V(r)]$ (see Table 2).

For the naked RN singularity, the quasinormal modes of the latter
type were calculated in the recent work \cite{Chirenti}. These
authors postulate the Dirichlet boundary condition that rules out
one of the linearly independent solutions to the Klein-Gordon
equation.  In the present work we do not specify the boundary
condition because it is not needed in the WKB approximation.
Moreover, for the metastable bound states we consider the point
$r=0$ is located deep into the classically forbidden region and thus
the boundary condition cannot affect these states substantially.

\begin{table}[pht]
\large \caption{Energies of the metastable bound states on the naked
RN background calculated with $Q=1.01$, $M=1$ and $m=0.1$.}
\begin{center}
{\begin{tabular}{cccc} \hline
& \multicolumn{3}{c}{ $l=1$} \\
\cline{2-4}
$\phantom{\displaystyle{\sum}}$  & $\omega_{WKB}$ & $\sqrt{V_{min}}$ & $\sqrt{V_{max}}$ \\
\hline
 $n=0$ & $0.295156-i\,0.221\times 10^{-3}$ &  0.241591 & 0.383910 \\
\hline
& \multicolumn{3}{c}{ $l=2$} \\
\cline{2-4}
 $\phantom{\displaystyle{\sum}}$ & $\omega_{WKB}$ & $\sqrt{V_{min}}$ & $\sqrt{V_{max}}$ \\
\hline
 $n=0$ & $0.423496-i\,0.142\times 10^{-5}$ & 0.364614 & 0.634529 \\
 $n=1$ & $0.537417-i\,0.209\times 10^{-3}$ &  &  \\
\hline
 & \multicolumn{3}{c}{ $l=3$} \\
\cline{2-4}
$\phantom{\displaystyle{\sum}}$   & $\omega_{WKB}$ & $\sqrt{V_{min}}$ & $\sqrt{V_{max}}$ \\
\hline
 $n=0$ & $0.555216-i\,0.869\times 10^{-10}$ & 0.494299 & 0.886277 \\
 $n=1$ & $0.673975-i\,0.219\times 10^{-5}$ &  &  \\
 $n=2$ & $0.786450-i\,0.200\times 10^{-3}$ &  &  \\
 $n=3$ & $0.885377-i\,0.569\times 10^{-2}$ &  &  \\
\hline
\end{tabular}}\end{center}
\label{ta3}
\end{table}

Next we investigate the case of the massive particle. In Table~3 the
energies of all the states with $l=1, 2, 3$ computed using $Q=1.01$,
$M=1$ and $m=0.1$ are listed. Also, in Fig.~4 the imaginary part of
$\omega$ is plotted versus the real part of $\omega$ for different
values of the particle mass $m$. From Tables 1, 3 and Fig.~4, we see
that the real part of $\omega$, \textit{i.e.} the oscillation
frequency, grows with increase of $m$, while the imaginary part of
$\omega$, representing the decay rate, falls down. Interestingly,
the same behavior was observed for the quasinormal modes of the
scalar field on the sub-extreme RN black-hole background in
Ref.~\cite{Konoplya}.

\begin{figure}[pht]
\begin{tabular}{cc}
 \includegraphics[scale=0.3]{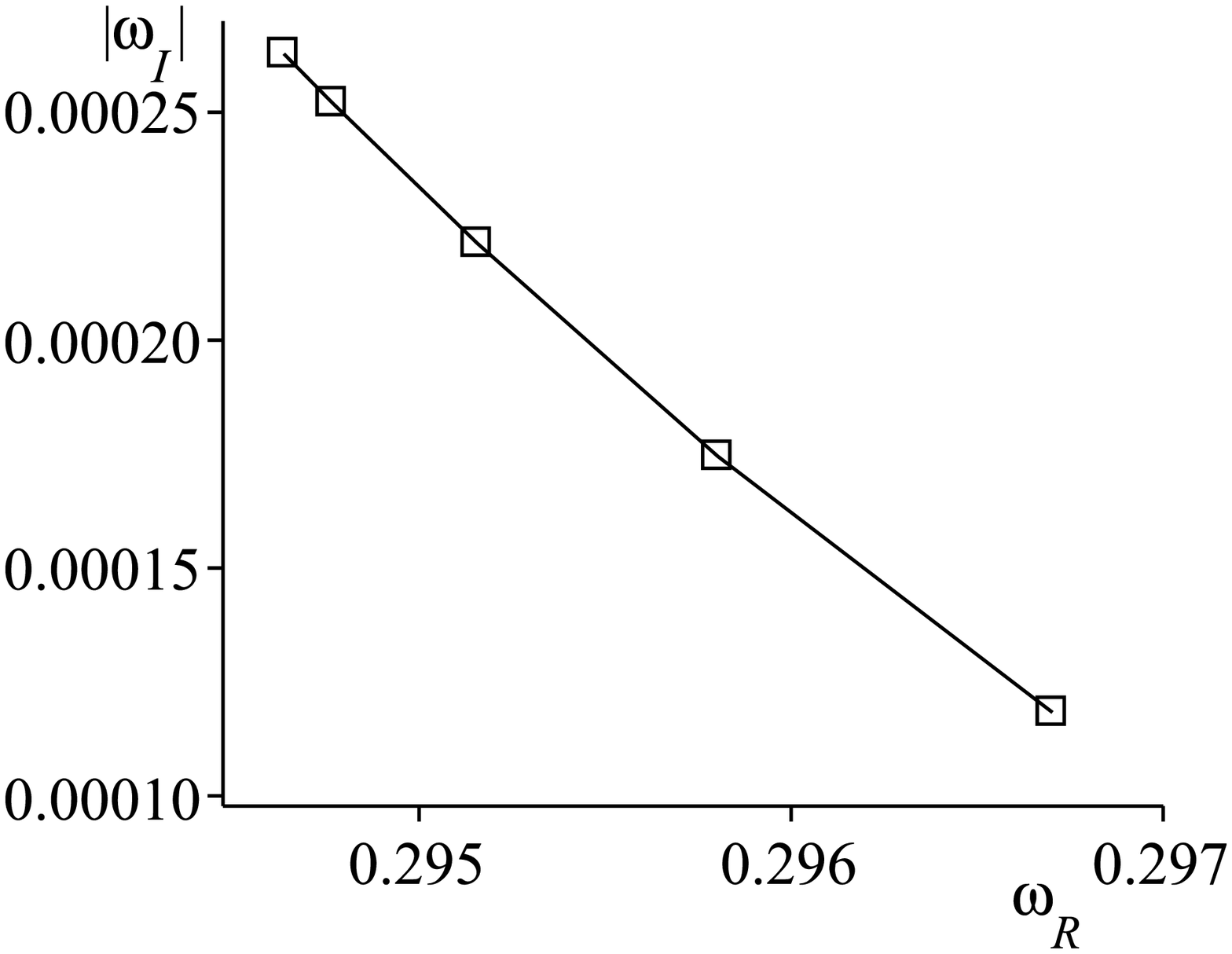} &  \includegraphics[scale=0.3]{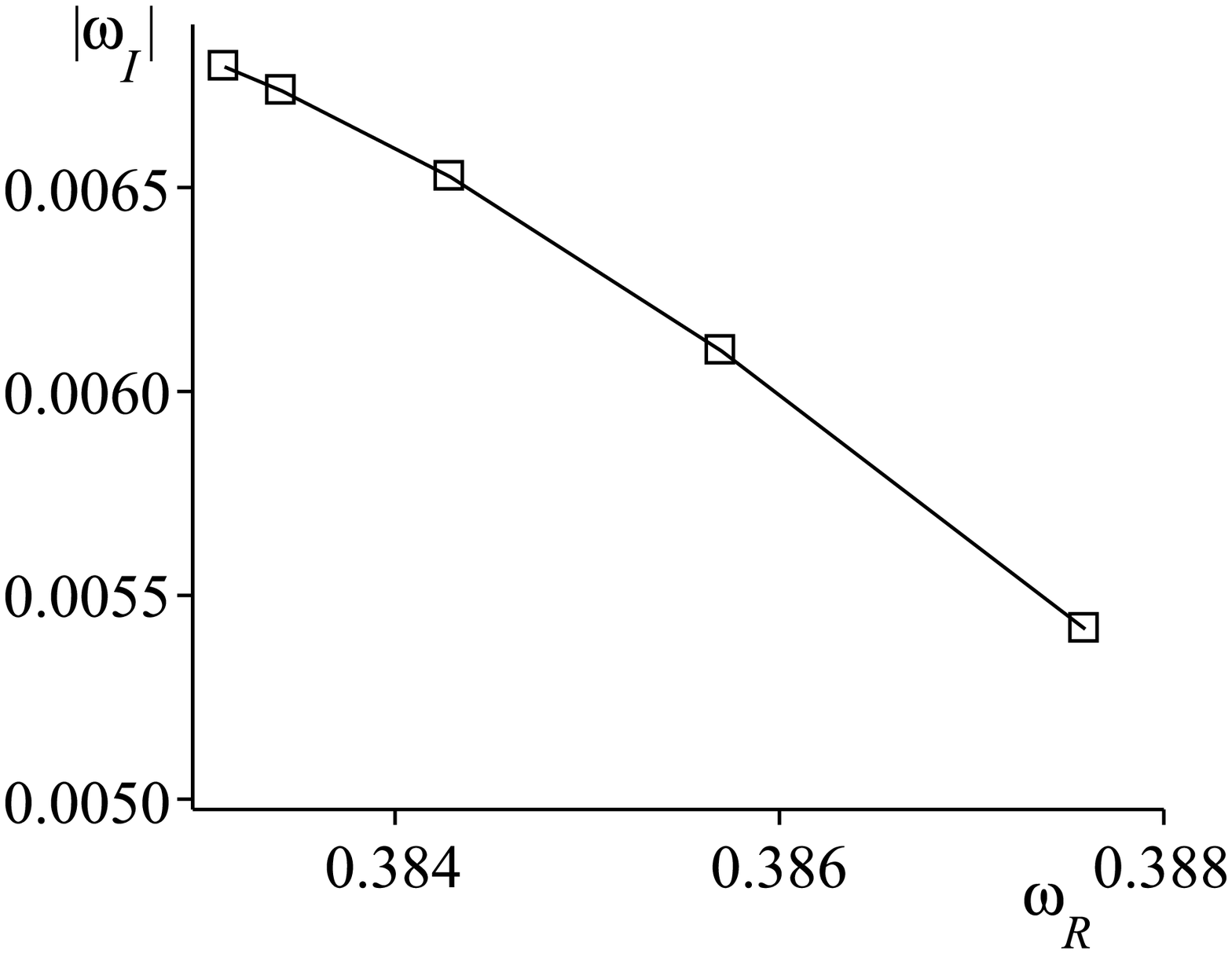} \\
  (a) & (b) \\
\end{tabular}
\label{fig4} \small \caption{ Real and imaginary parts of the
metastable bound-state energies for $n=0$, $l=1$ using (a) $M=1$,
$Q=1.01$, (b) $M=1$, $Q=1.02$. The squares left to right present the
results obtained with $m=0,0.05,0.1,0.15,0.2$ respectively. \hfill}
\end{figure}

Now let analyze the spectrum of the stable bound states. In Table~4
we list the energies $(\omega/m)_{WKB}$ of the ground and the first
two excited S-states ($l = 0$) computed according to the WKB formula
(\ref{WKBenergies}) and also the energies $(\omega/m)_{num}$
obtained by direct numerical integration of the Klein-Gordon
equation (\ref{RadEq}) with the Dirichlet boundary condition
$u(0)=0$. The calculation was made using $Q=1.1$, $M=1$ and various
$m$. From Table~4 we see that the binding energy
$\omega_{bind}=-(\omega-m)$ increases as the particle mass $m$
increases. However, this binding energy is more than two orders of
magnitude smaller than the metastable bound-state energy $\omega_R$
calculated with the same values of $m$ and $M$ (see Table 3). It
means that transitions between levels of the stable states due to an
external perturbation would have much lower frequencies than the
those due to the decay of the metastable bound states.

\begin{table}[pht]
\large \caption{Energies of the stable bound S-states on the naked
RN background with $Q=1.1$, $M=1$.}
\begin{center}
{\begin{tabular}{ccc|cc|cc} \hline
 & \multicolumn{2}{c|}{ $m=0.05$ } & \multicolumn{2}{c|}{ $m=0.1$ } & \multicolumn{2}{c}{ $m=0.15$ } \\
 \cline{2-7}
\vspace{1mm}
$\phantom{n=0}$ & $(\omega/m)_{WKB}$ & $(\omega/m)_{num}$ &  $(\omega/m)_{WKB}$ & $(\omega/m)_{num}$ &  $(\omega/m)_{WKB}$ & $(\omega/m)_{num}$ \\
 \cline{2-7}
$n=0$ & 0.998757 & 0.998733 &    0.995105 & 0.994710 &    0.989266 & 0.987068 \\
\hline
$n=1$ & 0.999688 & 0.998685 &    0.998764 & 0.998701 &    0.997257 & 0.996906 \\
\hline
$n=2$ & 0.999861 & 0.999860 &    0.999449 & 0.999429 &    0.998771 & 0.998661 \\
\hline
\end{tabular}}\end{center}
\label{ta4}
\end{table}

\section{Conclusion}
\label{part6}

In this work we have studied the states of neutral scalar particles
in the field of the naked RN singularity. It has been established
that their energy spectrum is qualitatively different from the
quasinormal mode spectrum for the RN black hole. The conditions for
the bound states to be formed have been found and the possible types
of these states have been examined.

The first possible type is the metastable, \textit{i.e.} decaying
bound states. They occur provided that the charge-to-mass ratio for
the central object, $Q/M$, is close to its extreme value $Q/M=1$. If
the particle is massless, this is expressed by the inequality
$1<(Q/M)^2<[8+9l(l+1)]/[8+8l(l+1)]$  where $l$ is the orbital
number.

Using the WKB method, we have calculated the metastable bound-state
spectrum which shows that there is no continuous transition in
energies between the RN naked singularity and the extremal RN black
hole cases. The metastable bound states on the naked RN background
with $Q/M>1$ have much larger live-times than the quasinormal modes
for $Q/M=1$. This is not surprising because for $Q/M>1$ there exists
no event horizon to fall on, so that the metastable states can decay
only at the expense of the particle escape to spatial infinity upon
tunneling through the wide potential barrier.

As the second possible type of particle states, we identify the
stable bound states. It should be stressed that these states have no
analog in the case of the RN black hole and can only be formed by
massive particles in the field of the central object with the
sufficiently high charge-to-mass ratio. Their energy spectrum,
calculated by means of the WKB method, has the structure similar to
that of the Coulomb spectrum.

Energy gaps between the stable bound states, \textit{i.e.}
transition frequencies, are several orders of magnitude smaller than
the frequencies for the metastable bound states obtained with the
same masses of the particle and the central object. Thus the latter
states are more favored to be ever found in experiment.
Nevertheless, the stable bound states may manifest themselves in a
different way, by making up a scalar condensate around the central
object. Then the question arises as to whether such a condensate can
provide mass needed to screen the RN singularity. Answering this
question requires a self-consistent model for interacting scalar and
gravitational fields which may be constructed in a future work.

\section*{Acknowledgments}
We thank the anonymous referee for a useful suggestion aimed at
improving the paper. This work was supported by the grant under the
Cosmomicrophysics program for the Physics and Astronomy Division of
the National Academy of Sciences of Ukraine.


\begin{thebibliography}{00}

\bibitem{Chandrasekhar} S. Chandrasekhar, \textit{The mathematical theory of black holes}, Clarendon Press, Oxford (1983).

\bibitem{Cohen} J. M. Cohen, R. Gautreau, {\it Phys. Rev. D}
\textbf{19} (1979) 2273.

\bibitem{Wald} R. M.  Wald {\it J. Math. Phys.}
\textbf{21} (1980) 2802.

\bibitem{Horowitz} G. T. Horowitz, D. Marolf, {\it Phys. Rev. D} \textbf{52} (1995) 5670.

\bibitem{Ishibashi99} A. Ishibashi, A. Hosoya, {\it Phys. Rev. D}
\textbf{60} (1999) 104028.

\bibitem{Pitelli} J. P. M. Pitelli, P. S. Letelier, {\it Int. J. Mod. Phys. D }
\textbf{20} (2011) 729.


\bibitem{Deruelle} N. Deruelle, R. Ruffini {\it Phys. Lett.}
\textbf{52B} (1974) 437.

\bibitem{Kokkotas} K. D.  Kokkotas, B. G. Schmidt {\it Living Rev. Rel.}
\textbf{2} (1999) 2. 

\bibitem{Berti} E. Berti, V. Cardoso {\it Class. Quant.
Grav.} \textbf{26} (2009) 163001. 



\bibitem{Qadir} A. Qadir, A. A. Siddiqui, {\it Int. J. Mod. Phys. D}
\textbf{16} (2007) 25.

\bibitem{Gladush} V.D. Gladush, M.V. Galadgyi, {\it Gen. Rel. Grav.}
\textbf{43} (2011) 1347.

\bibitem{Pugliese} D. Pugliese, H. Quevedo, R. Ruffini, {\it Phys. Rev. D}
\textbf{83} (2011) 024021.

\bibitem{Penrose} R. Penrose, {\it Ann. N. Y. Acad. Sci.}
\textbf{224} (1973) 125.

\bibitem{Damour} T. Damour, N. Deruelle, {\it Phys. Lett.}
\textbf{72B} (1978) 471.



\bibitem{Markov} M. A. Markov, V. P. Frolov, {\it Teor. Mat. Fiz.}
\textbf{3} (1970) 3.

\bibitem{Dirac} P. A. M. Dirac, {\it Proc. Roy. Soc. A}
\textbf{268} (1962) 57.




\bibitem{Ishibashi03} A. Ishibashi, R. M. Wald, {\it  Class. Quant. Grav.}
\textbf{20} (2003) 3815.

\bibitem{Stalker} J. G. Stalker, A. Shadi Tahvildar-Zadeh, {\it  Class. Quant. Grav.}
\textbf{21} (2004) 2831.

\bibitem{Kulikov} V. D. Gladush, D. A. Kulikov, arXiv:1110.3181 [gr-qc] (2011).

\bibitem{Ronveaux} A. Ronveaux (Ed.), {\it Heun's Differential Equations}, (Oxford University Press, Oxford, 1995).



\bibitem{Landau} L. D. Landau, E. M. Lifshitz, \textit{Quantum Mechanics. Nonrelativistic Theory}, Pergamon Press, Oxford (1965).

\bibitem{Heading} J. Heading, \textit{ An Introduction to Phase-Integral Methods}, Methhuen,
London; John Wiley, New York, (1962).

\bibitem{Berry} M. V. Berry, K. E. Mount, {\it Rep. Prog. Phys.}
\textbf{35} (1972) 315.

\bibitem{Mur} V. D. Mur, V. S. Popov, D. N. Voskresensky, {\it JETP Lett.}
\textbf{28} (1978) 129.

\bibitem{Onozawa} H. Onozawa, T. Mishima, T. Okamura, H. Ishihara, {\it Phys. Rev. D}
\textbf{53} (1996) 7033.

\bibitem{Chirenti} C. Chirenti,  A. Saa, J. Sk\'akala, {\it Phys. Rev. D}
\textbf{86} (2012) 124008.

\bibitem{Konoplya} R. A. Konoplya, {\it Phys. Lett. B}
\textbf{550} (2002) 117.


\end{thebibliography}
\end{document}